\documentclass[preprintnumbers, prd, twocolumn, showpacs, floatfix, preprintnumbers, 
letterpaper, 
superscriptaddress,nofootinbib]{revtex4-1}
\usepackage{amsmath}
\usepackage{amsfonts}
\usepackage{amssymb}
\usepackage{mathtools}
\usepackage{bm}
\usepackage{xcolor}
\usepackage{hyperref}
\usepackage{orcidlink}
\usepackage{graphicx}

\usepackage{color}
\usepackage{relsize}

\newcommand{\be}{\begin{equation}}
\newcommand{\ee}{\end{equation}}
\newcommand{\bea}{\begin{eqnarray}}
\newcommand{\eea}{\end{eqnarray}}

\begin{document}

\title{Phase-space analysis of torsion-coupled dilatonic ghost condensate}

\author{Manuel Gonzalez-Espinoza\orcidlink{0000-0003-0961-8029}}
\email{manuel.gonzalez@upla.cl}
\affiliation{Laboratorio de Did\'actica de la  F\'isica, Departamento de Matem\'atica\text{,} F\'isica y Computaci\'on, Facultad de Ciencias Naturales y Exactas, Universidad de Playa Ancha, Subida Leopoldo Carvallo 270, Valpara\'iso, Chile}
\affiliation{Laboratorio de investigación de Cómputo de Física, Facultad de Ciencias Naturales y
Exactas, Universidad de Playa Ancha, Subida Leopoldo Carvallo 270, Valparaíso, Chile}

\author{Giovanni Otalora\orcidlink{0000-0001-6753-0565}}
\email{giovanni.otalora@academicos.uta.cl}
\affiliation{Departamento de F\'isica, Facultad de Ciencias, Universidad de Tarapac\'a, Casilla 7-D, Arica, Chile}

\author{Yoelsy Leyva\orcidlink{0000-0002-4333-839X}}
\email{yoelsy.leyva@academicos.uta.cl }
\affiliation{Departamento de F\'isica, Facultad de Ciencias, Universidad de Tarapac\'a, Casilla 7-D, Arica, Chile}

\author{Joel Saavedra\orcidlink{0000-0002-1430-3008}}
\email{joel.saavedra@pucv.cl}
\affiliation{Instituto de F\'{\i}sica, Pontificia Universidad Cat\'olica de 
Valpara\'{\i}so, 
Casilla 4950, Valpara\'{\i}so, Chile}

\date{\today}

\begin{abstract} 
We studied the cosmological dynamics of a dilatonic ghost condensate field as a source of dark energy, which is non-minimally coupled to gravity through torsion. We performed a detailed phase-space analysis by finding all the critical points and their stability conditions. Also, we compared our results with the latest $H(z)$ and Supernovae Ia observational data. In particular, we found the conditions for the existence of scaling regimes during the dark matter era. Furthermore, we obtained the conditions for a successful exit from the scaling regime, such that, at late times, the universe tends towards an attractor point describing the dark energy-dominated era. These intriguing features can allow us to alleviate the energy scale problem of dark energy since, during a scaling regime, the field energy density is not necessarily negligible at early times.   

\end{abstract}


\maketitle

\section{Introduction}\label{Introduction}

The nature of dark energy, the mysterious ``force'' responsible for the observed accelerated expansion of the universe, is one of the biggest unsolved problems in cosmology. Despite decades of intense research, the origin and properties of dark energy remain elusive. One possible explanation for dark energy is provided by a dynamical scalar field. This idea has been extensively studied in the literature, for instance, quintessence \cite{Wetterich:1987fm,Ratra:1987rm,Carroll:1998zi,Tsujikawa:2013fta}, k-essence \cite{Chiba:1999ka,ArmendarizPicon:2000dh,ArmendarizPicon:2000ah}, scalar fields coupled to curvature or torsion or matter \cite{2004PhRvD..70l3518B,2004PhRvL..93q1104K,2007PhRvD..76f4004H,PhysRevD.77.046009, Lopez:2021agu,Gonzalez-Espinoza:2021qnv,Gonzalez-Espinoza:2021mwr,Gonzalez-Espinoza:2020jss,Gonzalez-Espinoza:2020azh,Gonzalez-Espinoza:2019ajd,Otalora:2014aoa,Otalora:2013dsa,Otalora:2013tba}, scalar fields with higher order derivatives in the action (also known as Galileons) \cite{Nicolis:2008in,Deffayet:2009wt,Baker:2017hug,Sakstein:2017xjx}, scalar-vector theories \cite{DeFelice:2016yws,DeFelice:2020icf,Armendariz-Picon:2004say,Koivisto:2008xf,Gonzalez-Espinoza:2022hui}, scalar fields in modified gravity with rainbow effects \cite{Leyva:2021fuo}, etc. However, the question of what is the exact nature of dark energy remains open.

On the other hand, another class of scalar fields has emerged as a potential source of dark energy: the so-called ghost condensate fields \cite{Arkani-Hamed:2003pdi,Arkani-Hamed:2003juy,Singh:2003vx,Copeland:2006wr}. These fields are characterized by a negative kinetic term, which in principle could give rise to quantum instabilities related to the presence of negative energy ghost states and lead to an unstable vacuum with a catastrophic production of ghost and photons pairs \cite{Copeland:2006wr, q-instabilities1}. Nevertheless, in the presence of higher-order kinetic terms, the field can "condensate" at a non-constant background value, providing a way to avoid these quantum instabilities. Ghost condensate fields have been extensively studied in the literature from a theoretical and observational perspective \cite{Arkani-Hamed:2003pdi,Arkani-Hamed:2003juy,Singh:2003vx,Copeland:2006wr,Creminelli:2005qk,Hussain:2022osn,Koehn:2012te,Peirone:2019aua}. The properties of ghost condensate fields make them an attractive candidate for dark energy, as they can drive the accelerated expansion of the universe while avoiding the problem of a cosmological constant. However, if a ghost condensate field describes dark energy, the stability of the quantum fluctuations cannot be guaranteed as the higher-order kinetic terms are suppressed due to the small energy density of the field relative to the Planck density \cite{Copeland:2006wr}. A possible solution to this problem is provided by the dilatonic ghost condensate model, which is motivated by dilatonic higher-order corrections to the tree-level action in low energy effective string theory \cite{Piazza:2004df}. The kinetic higher-order correction is now coupled to the dilaton field so that it is no longer suppressed at late times, allowing the stability of the vacuum while also generating cosmological solutions with accelerated expansion \cite{Copeland:2006wr}.   

Thus, the study of dilatonic ghost condensate field as a source of dark energy has attracted significant attention in recent years \cite{Piazza:2004df,Gumjudpai:2005ry,Peirone:2019aua}. Several theoretical models have been proposed to describe the behavior of the field, including models that incorporate the field in modified gravity theories such as $f(R)$ gravity and scalar-tensor theories \cite{DeFelice:2010aj,Horndeski:1974wa,Albuquerque:2018ymr,Kobayashi:2019hrl,Frusciante:2018vht}. These models have been tested against observational data, including data from the Cosmic Microwave Background (CMB), Large-Scale Structure Surveys (LSS), and type Ia Supernovae  (SNe Ia) \cite{Tsujikawa:2010sc}.

Despite the promising results, studying the dilatonic ghost condensate field as a source of dark energy is still in its early stages, and many questions remain unanswered. One of the biggest challenges in this field is to develop a consistent and viable theory that can explain the origin and behavior of the field. Another challenge is to develop observational tests that can distinguish between different models and constrain the parameters of the field.

So, the study of the dilatonic ghost condensate field as a source of dark energy is an exciting and promising field of research. The non-minimal coupling of the field to gravity through torsion leads to a unique behavior that can drive the accelerated expansion of the universe without the need for dark energy or a cosmological constant. These torsional modified gravity theories are constructed as extensions of the so-called Teleparallel Equivalent of General Relativity, or just Teleparallel Gravity (TG) \cite{Einstein,TranslationEinstein,Early-papers1,Early-papers2,Early-papers3,Early-papers4,Early-papers5,Early-papers6,JGPereira2,AndradeGuillenPereira-00,Arcos:2005ec,Pereira:2019woq,Aldrovandi-Pereira-book}. The best-known example of torsional modified gravity is $f(T)$ gravity \cite{Bengochea:2008gz,Linder:2010py,Li:2011wu}. This latter theory has also been extended to the case of generalized scalar-torsion $f(T,\phi)$ gravity theories to include a scalar field and its non-minimal interactions with gravity \cite{Hohmann:2018rwf,Gonzalez-Espinoza:2020azh,Gonzalez-Espinoza:2021qnv,Gonzalez-Espinoza:2021mwr,Gonzalez-Espinoza:2020jss}.
These kinds of modified gravity theories have proven to exhibit interesting features in the context of gravitation and cosmology \cite{Cai:2015emx}. Additionally, they have also demonstrated efficiency in fitting observations to solve both $H_{0}$ and $\sigma_{8}$ tensions simultaneously \cite{Yan:2019gbw}. The development of viable theories and observational tests in this field can provide important insights into the nature and origin of dark energy. It could have profound implications for our understanding of the universe.

This paper aims to investigate the cosmological behavior of a dilatonic ghost condensate field as source of dark energy in the framework of Modified Teleparallel Gravity. The paper is organized as follows: In Section \ref{Intro_TG}, we provide a brief introduction to the basics of Teleparallel Gravity. In Section \ref{model}, we present the total action of the model and the corresponding field equations. In Section \ref{cosmo_dyna}, we derive the cosmological equations and define the effective dark energy. In Section \ref{phase_space}, we perform a phase-space analysis of the model, including the autonomous system, the critical points, and stability conditions. In Section \ref{Num_Res}, we present numerical solutions to contrast with observational data. Finally, in Section \ref{conclusion_f}, we summarize the main results of the paper.


\section{A very short introduction to teleparallel gravity}\label{Intro_TG}
In Teleparallel Gravity (TG), a gauge theory for the translation group \cite{Aldrovandi-Pereira-book,JGPereira2,Arcos:2005ec},  the gravitational field is represented through torsion and not curvature \cite{Early-papers5,Early-papers6,Aldrovandi-Pereira-book,Pereira:2019woq}. Thus, the dynamical variable is the tetrad field which is related to the spacetime metric via
\be
g_{\mu \nu}=\eta_{A B} e^{A}_{~\mu} e^{B}_{~\nu}, 
\ee where $\eta _{AB}^{}=\text{diag}\,(-1,1,1,1)$ is the Minkowski tangent space metric. 

The Lorentz (or spin) connection of TG is defined as
\be
\omega^{A}_{~B \mu}=\Lambda^{A}_{~D}(x) \partial_{\mu}{\Lambda_{B}^{~D}(x)},
\label{spin_TG}
\ee being $\Lambda^{A}_{~D}(x)$ the components of a local (point-dependent) Lorentz transformation. This connection has a vanishing curvature tensor 
\be
R^{A}_{~B \mu\nu}=\partial_{\mu} \omega^{A}_{~B\nu}-\partial_{\nu}{\omega^{A}_{~B \mu}}+\omega^{A}_{~C \mu} \omega^{C}_{~B \nu}-\omega^{A}_{~C \nu} \omega^{C}_{~B \mu}=0, 
\ee which is the reason because this connection is also called a purely inertial connection or simply flat connection. On the other hand, in the presence of gravity this connection has a non-vanishing torsion tensor
\be
T^{A}_{~~\mu \nu}=\partial_{\mu}e^{A}_{~\nu} -\partial_{\nu}e^{A}_{~\mu}+\omega^{A}_{~B\mu}\,e^{B}_{~\nu}
 -\omega^{A}_{~B\nu}\,e^{B}_{~\mu}.
\ee 
Moreover, we can construct a spacetime-indexed linear connection which is given by
\be
\Gamma^{\rho}_{~~\nu \mu}=e_{A}^{~\rho}\partial_{\mu}e^{A}_{~\nu}+e_{A}^{~\rho}\omega^{A}_{~B \mu} e^{B}_{~\nu}.
\ee This is the popularly called Weitzenb\"{o}ck connection. Thus, using the contortion tensor 
\begin{equation}  \label{Contortion}
 K^{\rho}_{~~\nu\mu}= \frac{1}{2}\left(T^{~\rho}_{\nu~\mu}
 +T^{~\rho}_{\mu~\nu}-T^{\rho}_{~~\nu\mu}\right),
\end{equation} and the purely spacetime form of the torsion tensor
$T^{\rho}_{~~\mu \nu}=e_{A}^{~\rho} T^{A}_{~~\mu \nu}$
one can show that 
\be
\Gamma^{\rho}_{~~\nu \mu}=\bar{\Gamma}^{\rho}_{~~\nu \mu}+K^{\rho}_{~~\nu \mu},
\label{RelGamma}
\ee where $\bar{\Gamma}^{\rho}_{~~\nu \mu}$ is the known Levi-Civita connection of GR,  and such that 
\be
T^{\rho}_{~~\mu \nu}=\Gamma^{\rho}_{~~\nu \mu}-\Gamma^{\rho}_{~~\mu \nu}.
\ee

As a gauge theory, the action of TG is constructed using quadratic terms in the torsion tensor as
 \cite{Aldrovandi-Pereira-book}
\be
S=-\frac{1}{2 \kappa^2} \int{d^{4}x e ~T},
\ee where $e=\det{(e^{A}_{~\mu})}=\sqrt{-g}$ and $T$ is the torsion scalar such that
\be
T= S_{\rho}^{~~\mu\nu}\,T^{\rho}_{~~\mu\nu},
\label{ScalarT}
 \ee
with
\begin{equation} \label{Superpotential}
 S_{\rho}^{~~\mu\nu}=\frac{1}{2}\left(K^{\mu\nu}_{~~~\rho}+\delta^{\mu}_{~\rho} \,T^{\theta\nu}_{~~~\theta}-\delta^{\nu}_{~\rho}\,T^{\theta\mu}_{~~~\theta}\right)\,,
\end{equation} the super-potential tensor.  

By substituting Eq. \eqref{RelGamma} into \eqref{ScalarT} 
it is straightforward to verify that the torsion scalar of TG and curvature scalar of GR are related up to a total derivative term  
\be
T=-R+2 e^{-1} \partial_{\mu}(e T^{\nu \mu}_{~~~\nu}).
\label{Equiv} 
\ee Thus, one can observe that TG and GR are equivalent at the level of field equations. Moreover, when constructing modified gravity models, either the curvature-based or the torsion-based theory could be used as a starting point, and the resulting theories do not necessarily have to be equivalent. In the context of TG, dark energy and inflationary models driven by non-minimally coupled scalar field have been studied in Refs. \cite{Geng:2011aj,Otalora:2013tba,Otalora:2013dsa,Otalora:2014aoa,Skugoreva:2014ena}. On the other hand, modified gravity models constructed by adding to the gravitational action non-linear terms in torsion, as is the case of $f(T)$ gravity \cite{Bengochea:2008gz,Linder:2010py}, have also been studied in Refs. \cite{Li:2011wu,Gonzalez-Espinoza:2018gyl}. These torsion-based modified gravity theories belong to new classes of theories that do not have any curvature-based equivalent. Furthermore, a rich phenomenology has been found, which has given rise to a fair number of articles in the cosmology of early and late-time Universe \cite{Cai:2015emx}.

\section{Torsion-coupled dilatonic ghost condensate theory}\label{model}
We start with the action
\bea
S&=&-\int d^{4}{x} e\Bigg[\frac{T}{2\kappa^2}+F_1(\phi) T+X+V(\phi)- F_2(\phi) X^2 \Bigg]\nonumber\\
&& + S_{m}+ S_{r},
\eea 

where $X= -\frac{1}{2} \partial_{\mu}{\phi}\partial^{\mu}{\phi}$. $S_{m}$ is the action of matter. $S_{r}$ is the action of radiation. And varying the action with respect to the tetrad field, we obtain the field equations. Below we study the dynamics of the fields in a cosmological background. 

\section{Cosmological dynamics}\label{cosmo_dyna}
In order to study cosmology, we use the background tetrad field
\be
\label{veirbFRW}
e^A_{~\mu}={\rm
diag}(1,a,a,a),
\ee
which corresponds to a flat Friedmann-Lemaître-Robertson-Walker
(FLRW) universe with metric 
\begin{equation}
ds^2=-dt^2+a^2\,\delta_{ij} dx^i dx^j \,,
\label{FRWMetric}
\end{equation}
where $a$ is the scale factor which is a function of the cosmic time $t$. 
Therefore, the modified Friedmann equations are 
\bea
\label{MFreq1}
6 H^2 \left( \frac{1}{2 \kappa ^2} + F_1 \right)&=& V - \frac{\dot{\phi }^2}{2}  + \dfrac{3}{4} F_2 \dot{\phi}^4\nonumber\\
&+&\rho _m+\rho _r,\label{00}\\
-4 \dot{H} \left(\frac{1}{2 \kappa ^2} +  F_1\right)&=& - \dot{\phi }^2 + F_2 \dot{\phi}^4 + 4 H \dot{\phi } F_{1 ,\phi} \nonumber\\
&+&\rho _m+\frac{4 \rho _r}{3},\label{ii}
\eea along with the motion equation for $\phi$ is 
\bea
&&(3 F_2 \dot{\phi}^2 - 1)\ddot{\phi }+6 H^2 F_{1 ,\phi}+3 H \dot{\phi }(F_2 \dot{\phi}^2 -1 ) \nonumber\\
&&+V_{,\phi} + \dfrac{3}{4} F_{2 ,\phi} \dot{\phi}^4 =0.
\label{MFreq2}
\eea 

Thus, 

\bea
\label{SH00}
&& \frac{3}{\kappa^2} H^2=\rho_{de}+\rho_{m}+\rho_{r},\\
&& -\frac{2}{\kappa^2} \dot{H}=\rho_{de}+p_{de}+\rho_{m}+\frac{4}{3}\rho_{r},
\label{SHii}
\eea where the effective energy and pressure densities are defined as  
\bea
\label{rhode}
\rho_{de}&= & - \frac{\dot{\phi }^2}{2}  + \dfrac{3}{4} F_2 \dot{\phi}^4+V- 6 H^2 F_1,\\
 p_{de}&= & -\frac{\dot{\phi }^2}{2} - \dfrac{1}{4} F_2 \dot{\phi}^4-V + 6 H^2 F_{1}+4 H \dot{\phi } F_{1 ,\phi}+ 4 F_1 \dot{H}. \nonumber\\
 \label{pde}
&&\eea 
From Eqs. \eqref{MFreq1} and \eqref{MFreq2}, as well as from Eqs. \eqref{rhode} and \eqref{pde}, one can observe the emergence of a negative kinetic term. This negative kinetic term originates from the Lagrangian of the dilatonic ghost condensate field, which is a phantom (ghost) field that includes dilatonic higher-order corrections to guarantee the quantum stability of the vacuum \cite{Piazza:2004df, Copeland:2006wr}. In Appendix \ref{appen_A}, we have provided details on the required quantum stability conditions.   
Then, the effective dark energy equation-of-state (EOS) parameter is
\begin{equation}
w_{de}=\frac{p_{de}}{\rho _{de}}.
\label{wDE1}
\end{equation}
For these definitions of $\rho_{de}$ and $p_{de}$ one can verify that they satisfy
\begin{eqnarray}
\dot{\rho}_{de}+3H(\rho_{de}+p_{de})=0.
\end{eqnarray} This equation is consistent with the energy conservation law and the fluid evolution equations
\bea
\label{rho_m}
&& \dot{\rho}_{m}+3 H\rho_{m}=0,\\
&& \dot{\rho}_{r}+4 H\rho_{r}=0.
\label{rho_r}
\eea

Also, it is useful to introduce the
total EOS parameter as 
\be
w_{tot}=\frac{p_{de}+p_r}{\rho _{de}+\rho _m+\rho _r},
\label{wtot}
\ee which is related to the deceleration parameter $q$ through
\be
q=\frac{1}
{2}\left(1+3w_{tot}\right).
\label{deccelparam}
\ee Then, the acceleration of the Universe occurs for $q<0$, or equivalently for  $w_{tot}<-1/3$.

Finally, another useful set of cosmological parameters which we can introduce is the so-called standard density parameters  
\bea
&& \Omega_{m}\equiv\frac{\kappa^2 \rho_{m}}{3 H^2},\:\:\:\: \Omega_{de}\equiv\frac{\kappa^2 \rho_{de}}{3 H^2},\:\:\:\: \Omega_{r}\equiv \frac{\kappa^2 \rho_{r}}{3 H^2},
\eea which satisfies the constraint equation  
\be
\Omega_{de}+\Omega_{m}+\Omega_{r}=1.
\ee 
Below we perform a detailed dynamical analysis for this model. 

\section{Phase space Analysis}\label{phase_space}
We introduce the following useful dimensionless variables \cite{Copeland:2006wr}

\begin{table*}[ht]
 \centering
 \caption{Critical points for the autonomous system. We have defined $f_{\pm} (\sigma_2) = 1 \pm \sqrt{1+16/(3 \sigma_2)}$.}
\begin{center}
\begin{tabular}{c c c c c c c c c}\hline\hline
Name &  $x_c$ & $y_c$ & $u_{c 1}$ & $u_{c 2}$  & $\varrho_{c}$  \\\hline
$\ \ \ \ \ \ \ \ a_{R} \ \ \ \ \ \ \ \ $ & $0$ & $0$ & $0$ & $0$  & $1$ \\
$\ \ \ \ \ \ \ \ b_{M} \ \ \ \ \ \ \ \ $ & $0$ & $0$ & $0$ & $0$  & $0$ \\
$\ \ \ \ \ \ \ \ c_{M} \ \ \ \ \ \ \ \ $ & $ \sqrt{\dfrac{3}{2}}\dfrac{1}{\sigma_2}$ & $0$ & $0$ & $\sqrt{\dfrac{3}{2}}\dfrac{1}{\sigma_2}$  & $0$ \\
$\ \ \ \ \ \ \ \ d \ \ \ \ \ \ \ \ $ & $0$ & $\dfrac{\sqrt{\sigma _1}}{\sqrt{\lambda+\sigma _1}}$ & $\dfrac{\lambda}{\lambda+\sigma _1}$ & $u_{c2}$  & $0$ \\
$\ \ \ \ \ \ \ \ e \ \ \ \ \ \ \ \ $ & $-\dfrac{\sqrt{6} \sigma_2 f_+}{4}$ & $0$ & $0$ & $-\dfrac{\sqrt{6} \sigma_2 f_+}{\sqrt{8+\sigma_2^2 f_-}}$  & $0$ \\
$\ \ \ \ \ \ \ \ f \ \ \ \ \ \ \ \ $ & $-\dfrac{\sqrt{6} \sigma_2 f_-}{4}$ & $0$ & $0$  & $-\dfrac{\sqrt{6} \sigma_2 f_-}{\sqrt{8+\sigma_2^2 f_+}}$ & $0$\\

\\ \hline\hline
\end{tabular}
\end{center}
\label{table1}
\end{table*}
\begin{table}[ht]
 \centering
 \caption{Cosmological parameters for the critical points in Table \ref{table1}. }
\begin{center}
\begin{tabular}{c c c c c c}\hline\hline
Name &   $\Omega_{de}$ & $\Omega_{m}$ & $\Omega_{r}$ & $\omega_{de}$ & $\omega_{tot}$ \\\hline
$a_{R}$ & $0$ & $0$ & $1$ & $0$ & $\frac{1}{3}$ \\
$b_{M}$ & $0$ & $1$ & $0$ & $0$ & $0$ \\
$c_{M}$ & $\frac{3}{\sigma_2^2}$ & $1-\frac{3}{\sigma_2^2}$ & $0$ & $0$ & $0$ \\
$d$ & $1$ & $0$ & $0$ & $-1$ & $-1$ \\
$e$ & $1$ & $0$ & $0$ & $-1-\dfrac{\sigma_2^2 f_+}{2}$ & $-1-\dfrac{\sigma_2^2 f_+}{2}$ \\
$f$ & $1$ & $0$ & $0$ & $-1-\dfrac{\sigma_2^2 f_- }{2}$ & $-1-\dfrac{\sigma_2^2 f_- }{2}$ \\
\\ \hline\hline
\end{tabular}
\end{center}
\label{table2}
\end{table}

\begin{eqnarray}
x =& \dfrac{\kappa  \dot{\phi}}{\sqrt{6} H}, \ \ \ \ \ \ y =& \dfrac{\kappa  \sqrt{V}}{\sqrt{3} H}, \ \ \ \ \ \ \ u_1 = -2 \kappa ^2 F_{1},\nonumber\\ 
u_2 =& \dfrac{\kappa }{\sqrt{3 F_2} H}, \ \ \ \ \ \
\lambda =& - \dfrac{V_{,\phi}}{\kappa V}, \ \ \ \ \ \ \ \sigma_1 = - \dfrac{F_{1,\phi}}{\kappa F_1}, \nonumber\\
\sigma_2 =&  \dfrac{F_{2,\phi}}{\kappa F_2}, \ \ \ \ \ \ \ \Gamma_1 =&  \dfrac{V V_{,\phi\phi}}{(V_{,\phi})^2}, \ \ \ \ \ \ \ \Theta_1 = \dfrac{F_1 F_{1,\phi\phi}}{(F_{1,\phi})^2}, \nonumber\\
 \Theta_2 =& \dfrac{F_2 F_{2,\phi\phi}}{(F_{2,\phi})^2},\ \ \ \ \ \ \ \varrho= &\frac{\kappa\sqrt{\rho_r}}{\sqrt{3}H},\ \nonumber\\
&& 
\label{var}
\end{eqnarray}
and the constraint equation
\begin{equation}
    \Omega_m + \varrho^2 +\frac{3 x^4}{u_2^2}+u_1-x^2+y^2= 1,
\end{equation}
Therefore, we obtain the dynamical system
\begin{eqnarray}
\dfrac{d x}{d N} &=& \frac{f_1(x,y,u_1, u_2,\varrho)}{2 \left(u_1-1\right) u_2^2 \left(u_2^2-6 x^2\right)},
\label{dinsyseq1}\\
\dfrac{d y}{d N} &=& \frac{y f_2(x,y,u_1, u_2,\varrho)}{2 \left(u_1-1\right) u_2^2}, \label{dinsyseq2}\\
\dfrac{d u_1}{d N} &=& -\sqrt{6} \sigma _1 u_1 x, \label{dinsyseq3}\\
\dfrac{d u_2}{d N} &=& \frac{ f_3(x,y,u_1, u_2,\varrho)}{2 \left(u_1-1\right) u_2}, \label{dinsyseq4}\\
\dfrac{d \varrho}{d N} &=& -\frac{\varrho f_4(x,y,u_1, u_2,\varrho)}{2 \left(u_1-1\right) u_2^2}, \label{dinsyseq5}\\
\dfrac{d \lambda}{d N} &=& -\sqrt{6} \left(\Gamma _1-1\right) \lambda ^2 x, \label{dinsyseq6}\\
\dfrac{d \sigma_1}{d N} &=& -\sqrt{6} \left(\Theta _1-1\right) \sigma _1^2 x, \label{dinsyseq7}\\
\dfrac{d \sigma_2}{d N} &=& \sqrt{6} \left(\Theta _2-1\right) \sigma _2^2 x, \label{dinsyseq8}
\end{eqnarray}

where
{\small
\begin{eqnarray}
f_1 &=& u_2^4 \Big[\sqrt{6} \sigma _1 u_1^2-u_1 \left(\sqrt{6} \sigma _1 \left(2 x^2+1\right)+3 x+\sqrt{6} \lambda  y^2\right)+3 x^3 \nonumber\\
&+&x \left(3 y^2-\varrho ^2+3\right)+\sqrt{6} \lambda  y^2\Big]-3 u_2^2 x^3 \Big[-u_1 \Big(4 \sqrt{6} \sigma _1 x \nonumber\\
&+&\sqrt{6} \sigma _2 x-2\Big)+7 x^2+\sqrt{6} \sigma _2 x+6 y^2-2 \varrho ^2-2\Big]+18 x^7, \nonumber\\
f_2 &=& u_2^2 \Big[-u_1 \left(\sqrt{6} \lambda  x+2 \sqrt{6} \sigma _1 x-3\right)+3 x^2+\sqrt{6} \lambda  x+3 y^2 \nonumber\\
&-&\varrho ^2-3\Big]-3 x^4, \nonumber\\
f_3&=& u_2^2 \Big[-u_1 \left(2 \sqrt{6} \sigma _1 x+\sqrt{6} \sigma _2 x-3\right)+3 x^2+\sqrt{6} \sigma _2 x+3 y^2 \nonumber\\
&-&\varrho ^2-3\Big]-3 x^4, \nonumber\\
f_4 &=& u_2^2 \left[u_1 \left(2 \sqrt{6} \sigma _1 x+1\right)-3 x^2-3 y^2+\varrho ^2-1\right]+3 x^4,
\end{eqnarray}}

Using the above set of phase space variables, we can also write
\bea
\Omega_{de} &=& \frac{3 x^4}{u_2^2}+u_1-x^2+y^2,\\
\Omega_{m} &=& -\frac{3 x^4}{u_2^2}-u_1+x^2-y^2-\varrho ^2+1. \nonumber\\
&&
\eea

Similarly, the equation of state of dark energy $w_{de}=p_{de}/\rho_{de}$ can be rewritten as
\bea
w_{de} &=& -\frac{u_2^2 \left[u_1 \left(2 \sqrt{6} \sigma _1 x+\varrho ^2\right)-3 \left(x^2+y^2\right)\right]+3 x^4}{3 \left(u_1-1\right) u_2^2 \left(\frac{3 x^4}{u_2^2}+u_1-x^2+y^2\right)}, \nonumber\\
&&
\eea
whereas the total equation of state becomes
\bea
w_{tot}&=&\frac{u_2^2 \left(-2 \sqrt{6} \sigma _1 u_1 x+3 x^2+3 y^2-\varrho ^2\right)-3 x^4}{3 \left(u_1-1\right) u_2^2}. \nonumber\\
\eea

Finally, we need to define the potential, the non-minimal coupling to gravity and the coupling to $X^2$ to obtain an autonomous system from the dynamical system \eqref{dinsyseq1}-\eqref{dinsyseq8}. This paper uses exponential functions, $V = e^{- \kappa \lambda \phi}$, $F_1 = e^{- \kappa \sigma_1 \phi}$ and $F_2 = e^{\kappa \sigma_2 \phi}$ \cite{Copeland:2006wr}. The motivation behind the coupling, $F_2 = e^{\kappa \sigma_2 \phi}$, derives from including dilatonic higher-order corrections within the tree-level action of low-energy effective string theory \cite{Piazza:2004df}.

\subsection{Critical points}

The critical points are obtained by satisfying the conditions $dx/dN = dy/dN = du_1/dN=du_2/dN=d\varrho/dN=0$, and their values are real according to the definitions in Eq. \eqref{var}. Additionally, $y>0$, $u_{2}>0$, and $\varrho>0$ hold true for the critical points. Table \ref{table1} shows the critical points for the system \eqref{dinsyseq1}-\eqref{dinsyseq5}, and their corresponding values of cosmological parameters are presented in Table \ref{table2}.

Critical point $a_{R}$ describes a radiation-dominated era with $\Omega_{r}=1$. Critical point $b_{M}$ describes a matter-dominated era with $\Omega_{m}=1$ and $\omega_{de}=\omega_{tot}=0$. Also, the scaling matter era with $\Omega_{de}=3/\sigma_{2}^2$ is represented by the critical point $c_{M}$. Thus, the standard matter-dominated era is recovered for $\sigma_{2}\rightarrow \pm \infty$. It is important to note that the scaling solution $c_{M}$ is generated by the coupling function of the dilaton field in the higher-order kinetic term. This coupling function is specifically associated to the phase-space variable $u_{2}$, rather than the variable $y$, that in our analysis, is directly connected to the scalar potential \cite{Copeland:2006wr}. 

On the other hand, critical point $d$ is a dark energy-dominated solution with a de Sitter equation of state $\omega_{de}=\omega_{tot}=-1$, providing accelerated expansion. Points $e$ and $f$ are dark energy-dominated solutions, which receive contributions from the higher-order kinetic term $F_2(\phi) X^2$.

\subsection{Stability of critical points}\label{Stability}

To analyze the stability of critical points, in Table \ref{table2}, we assume time-dependent linear perturbations  $\delta x$, $\delta y$, $\delta u_1$, $\delta u_2$ and $ \delta\varrho$ around each critical point, in other words, we begin our analysis with the equations $x=x_c +\delta x$, $y=y_c+\delta y$, $u_1 = u_{1 c}+\delta u_1$, $u_2 = u_{2 c}+\delta u_2$ and $\varrho= \varrho_c + \delta\varrho$.  Then, after replacing these equations in the autonomous system \eqref{dinsyseq1}-\eqref{dinsyseq5} and considering only linear order terms, we obtain the linear perturbation matrix $\mathcal{M}$. Then, evaluating the eigenvalues, $\mu_1$, $\mu_2$, $\mu_3$, $\mu_4$ and $\mu_5$, of $\mathcal{M}$, for each critical point, we can determine the stability of these points. Where the stability properties are classified by: (i) stable mode: all eigenvalues are positive, (ii) unstable mode: all eigenvalues are negative, (iii) saddle point: at least one eigenvalue is positive, and the others are negative; and (iv) stable spiral: the real part of eigenvalues are positive and the determinant of $\mathcal{M}$ is negative \cite{Copeland:2006wr}. From these properties, and independently of initial conditions, we can determine the cosmological evolution of our Universe. \\

Subsequently, we analyze the stability of each critical point,

\begin{itemize}
    \item Point $a_R$ has the eigenvalues
    \begin{equation}
        \mu_1 = 2, \ \ \ \ \ \mu_2 = 2, \ \ \ \ \ \mu_3 = -1, \ \ \ \ \ \mu_4 = 1, \ \ \ \ \ \mu_5 = 0, 
    \end{equation}
    therefore, it is a saddle point.
    \item Point $b_M$ has the eigenvalues
    \begin{equation}
        \mu_1 = -\dfrac{3}{2}, \ \ \ \ \ \mu_2 = \dfrac{3}{2}, \ \ \ \ \ \mu_3 = \dfrac{3}{2}, \ \ \ \ \ \mu_4 = -\dfrac{1}{2}, \ \ \ \ \ \mu_5 = 0, 
    \end{equation}
    therefore, it is a saddle point.
    \item Point $c_M$ has the eigenvalues
    \begin{eqnarray}
        \mu_1 &=& -\frac{3 \sigma _1}{\sigma _2}, \ \ \ \ \ \mu_2 = -\frac{1}{2}, \ \ \ \ \ \mu_3 = \frac{3 \left(\sigma _2-\lambda\right)}{2 \sigma _2},\nonumber\\
        \mu_4 &=& -\frac{3 \sqrt{\frac{3}{5}} \sqrt{8-\sigma _2^2}}{4 \sigma _2}-\frac{3}{4}, \ \ \ \ \ \ \nonumber \\
        \mu_5 &=& \frac{3}{20} \left(\frac{\sqrt{15} \sqrt{8-\sigma _2^2}}{\sigma _2}-5\right),
    \end{eqnarray}
    where, if we consider $\Omega_{de}=\frac{3}{\sigma_2^2}<1$, this critical point is a saddle point for
    \begin{eqnarray}
        && \left(\sigma _1<0\land \sigma _2>\sqrt{3}\right)\lor \left(\sigma _1>0\land \sigma _2<-\sqrt{3}\right)\lor \nonumber\\
        && \left(\lambda\leq -\sqrt{3}\land \left(\sigma _2<\lambda\lor \sigma _2>\sqrt{3}\right)\right)\lor \nonumber\\
        && \left(-\sqrt{3}<\lambda<\sqrt{3}\land \left(\sigma _2<-\sqrt{3}\lor \sigma _2>\sqrt{3}\right)\right)\lor \nonumber\\
        && \left(\lambda\geq \sqrt{3}\land \left(\sigma _2<-\sqrt{3}\lor \sigma _2>\lambda\right)\right).
    \end{eqnarray}
    \item Point $d$ has the eigenvalues
    \begin{eqnarray}
        \mu_1 &=& -2, \ \ \ \ \ \mu_2 = -3, \ \ \ \ \ \mu_3 = 0,\nonumber\\
        \mu_4 &=& \frac{1}{2} \left(-\sqrt{3} \sqrt{3-4 \lambda \sigma_1}-3\right), \ \ \ \ \ \ \nonumber \\
        \mu_5 &=& \frac{1}{2} \left(\sqrt{3} \sqrt{3-4 \lambda \sigma_1}-3\right), \ \ \ \ \ \
    \end{eqnarray}    
    therefore, it is stable for
    \begin{equation}
        \left(\sigma_1<0\land \frac{3}{4 \sigma_1}\leq \lambda<0\right)\lor \left(\sigma_1>0\land 0<\lambda\leq \frac{3}{4 \sigma_1}\right).
    \end{equation}
    \item Point $e$ is stable for the following range of the values of parameters
    \begin{eqnarray}
        &&\left(\lambda \leq -\sqrt{3}\land -\sqrt{3}<\sigma _2<0\land \sigma _1<0\right)\nonumber\\
        &&\lor \left(-\sqrt{3}<\lambda <0\land \lambda <\sigma _2<0\land \sigma _1<0\right) \nonumber\\
        &&\lor \left(0<\lambda \leq \sqrt{3}\land 0<\sigma _2<\lambda \land \sigma _1>0\right)\nonumber\\
        &&\lor \left(\lambda >\sqrt{3}\land 0<\sigma _2<\sqrt{3}\land \sigma _1>0\right).
    \end{eqnarray}
    \item Point $f$ is stable for the following range of the values of parameters
    \begin{eqnarray}
        &&\left(\lambda \leq 0\land \left(\left(\sigma _2<\lambda \land \sigma _1>0\right)\lor \left(\sigma _2>0\land \sigma _1<0\right)\right)\right) \nonumber\\
        &&\lor \left(\lambda >0\land \left(\left(\sigma _2<0\land \sigma _1>0\right)\lor \left(\sigma _2>\lambda \land \sigma _1<0\right)\right)\right). \nonumber\\
        &&
    \end{eqnarray}
\end{itemize}

Below we corroborate our analytical results by numerically integrating the system of cosmological equations for our model. Then, we compare it with observational data. In Appendix \ref{appen_A}, we have also verified that our results satisfy the quantum stability conditions required for the dilatonic ghost condensate field \cite{Copeland:2006wr}.  

\begin{figure}[!tbp]
  \centering
  \begin{minipage}[b]{0.45\textwidth}
    \includegraphics[width=\textwidth, height=4cm]{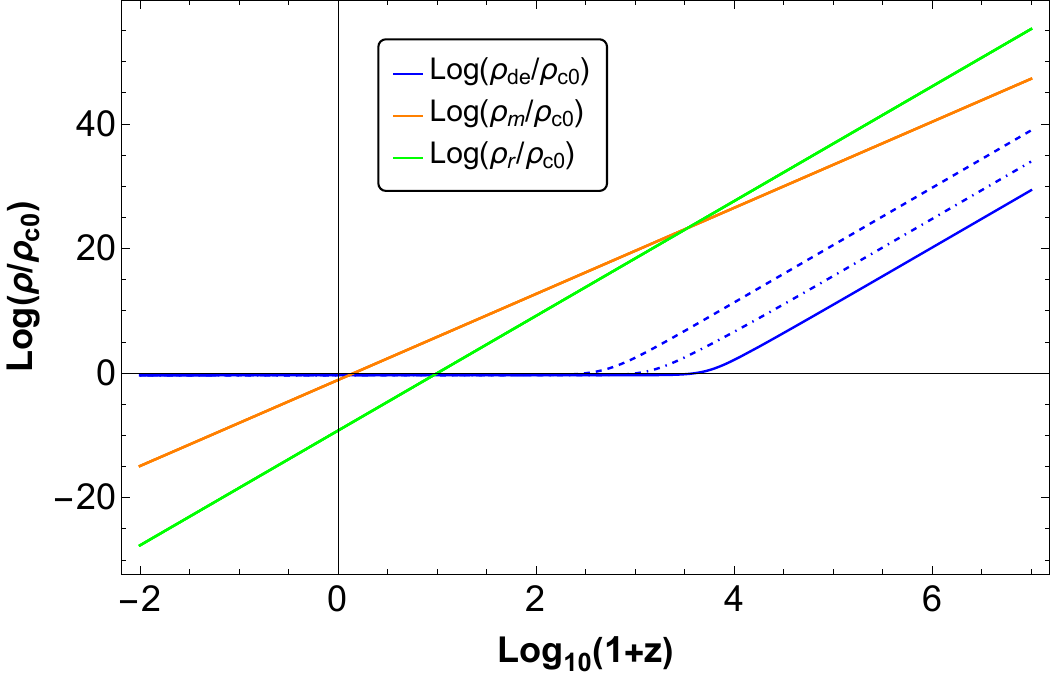}
    \caption{\scriptsize{We present the evolution of the energy density of dark energy, denoted as $\rho_{de}$ (depicted in blue), the energy density of dark matter (including baryons), denoted as $\rho_m$ (shown in orange), and the energy density of radiation, denoted as $\rho_r$ (illustrated in green), as functions of the redshift $z$. The values of the parameters used are $\lambda = 1$, $\sigma_1 = 1$, and $\sigma_2 = 10^{-3}$. The solid lines represent the cases with initial conditions $x_i = 1.7 \times 10^{-12}$, $y_i = 7.3 \times 10^{-14}$, $u_{1i} = 5.75 \times 10^{-12}$, $u_{2i} = 1.7 \times 10^{-12}$, and $\varrho_i = 0.99983$. On the other hand, the dashed lines correspond to initial conditions $x_i = 1.7 \times 10^{-8}$, $y_i = 7.3 \times 10^{-14}$, $u_{1i} = 5.75 \times 10^{-11}$, $u_{2i} = 1.71 \times 10^{-12}$, and $\varrho_i = 0.99983 $. Lastly, the dot-dashed lines represent the cases with initial conditions $x_i = 1.7 \times 10^{-12}$, $y_i = 7.3 \times 10^{-14}$, $u_{1i} = 5.75 \times 10^{-10}$, $u_{2i} = 1.69 \times 10^{-12}$, and $\varrho_i = 0.99983$. During the scaling radiation era, denoted as $a_R$, we have obtained an approximate value of $\Omega_{de}^{(r)} \approx 8.7275 \times 10^{-8}$.\\
    }} 
    \label{Figura1rho}
  \end{minipage}
  \hfill
  \begin{minipage}[b]{0.45\textwidth}
    \includegraphics[width=\textwidth, height=4cm]{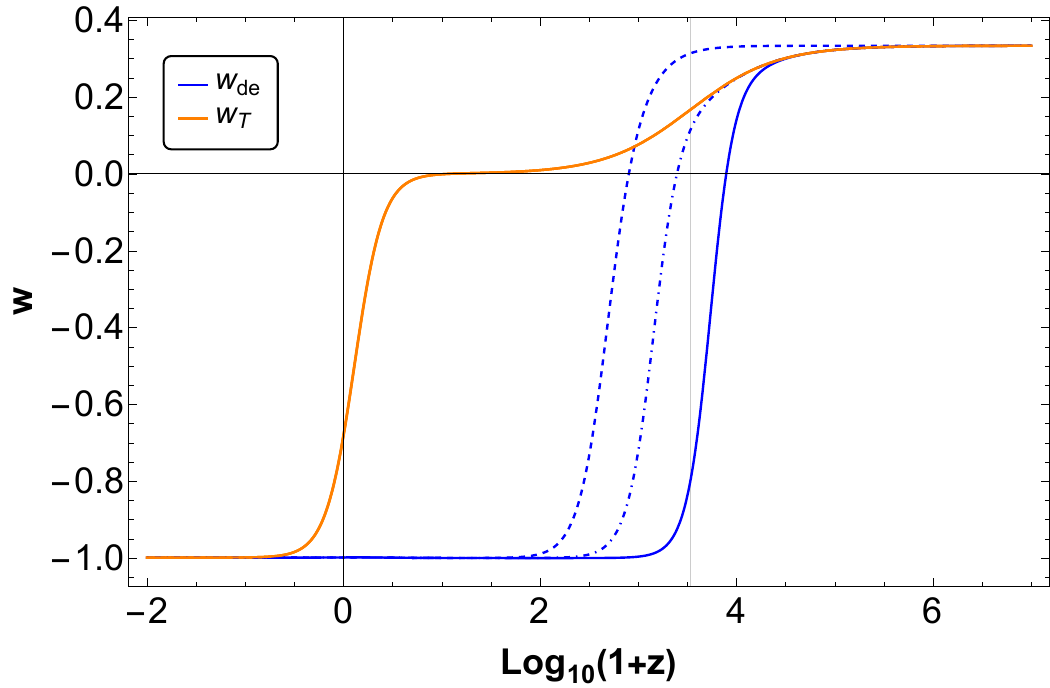}
    \caption{\scriptsize{The equation of state $w_{tot}$ (depicted by the orange line) and the dark energy equation of state $w_{de}$ (represented by the blue line) are plotted as functions of redshift $z$. The solid, dashed, and dot-dashed blue lines correspond to the initial conditions in FIG. \ref{Figura1rho}. At $z=0$, we obtain a value of $w_{de}=-0.998427$ for the dark energy equation of state. This value is in agreement with the Planck data value of $w_{de}^{(0)}=-1.028\pm0.032$ as reported in the Ref. \cite{Akrami:2018odb}.}} 
    \label{Figura1omega}
  \end{minipage}
\end{figure}

\begin{figure}[!tbp]
  \centering
  \begin{minipage}[b]{0.45\textwidth}
    \includegraphics[width=\textwidth, height=4cm]{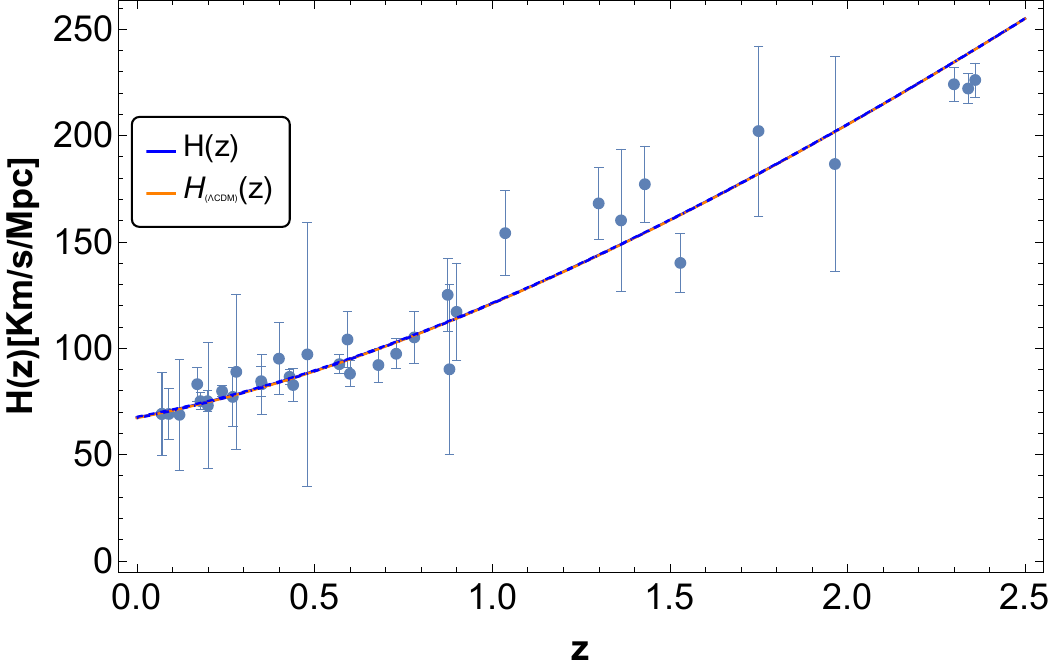}
    \caption{\scriptsize{We depict the evolution of the Hubble rate, denoted as $H(z)$, as a function of redshift $z$. The blue points represent the observed data points for the Hubble rate, and their corresponding $1 \sigma$ confidence intervals are provided in Table \ref{table:H(z)data}. The solid, dashed, and dot-dashed blue lines are obtained using the same initial conditions shown in FIG. \ref{Figura1rho}. By comparing the model predictions represented by the solid, dashed, and dot-dashed blue lines with the observed data, we can evaluate the agreement between the theoretical model and the empirical measurements of the Hubble rate at different redshifts.\\}} 
    \label{Figura1H}
  \end{minipage}
  \hfill
  \begin{minipage}[b]{0.45\textwidth}
    \includegraphics[width=\textwidth, height=4cm]{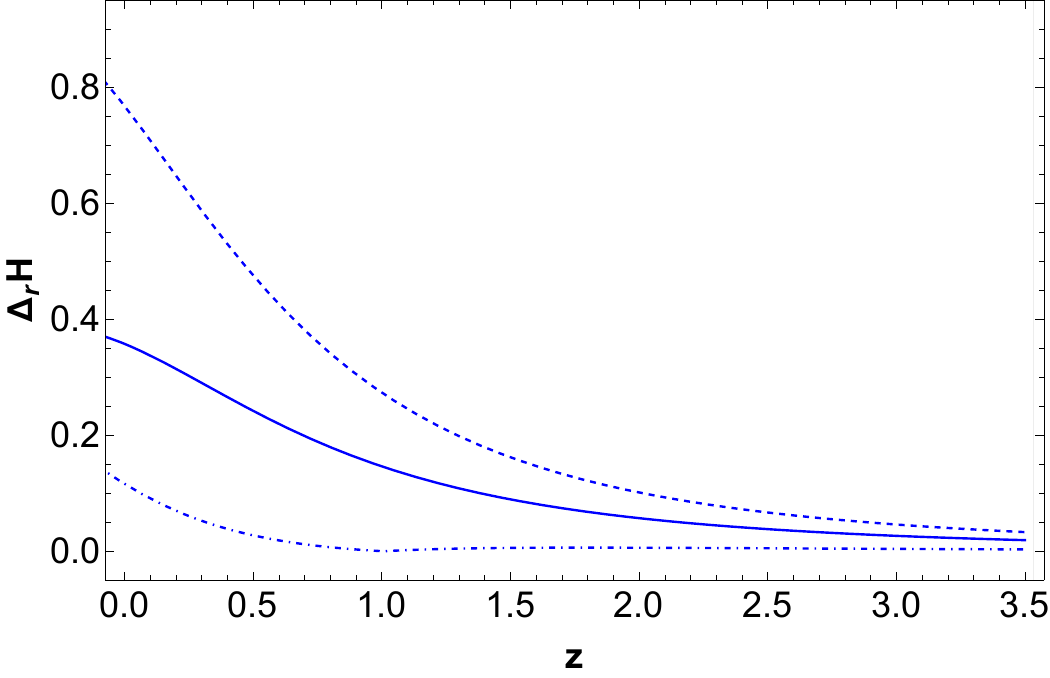}
    \caption{\scriptsize{We show the behavior of relative difference (percentage), $\Delta_r H(z) = 100 \times \left| H - H_{\Lambda CDM} \right|/H_{\Lambda CDM}$, with respect to $\Lambda CDM$ model where we used the same initial conditions of FIG. \ref{Figura1rho} to obtain the solid, dashed, and dot-dashed blue lines.}} 
    \label{FiguradH1}
  \end{minipage}
\end{figure}


\begin{figure}[!tbp]
	\centering
		\includegraphics[width=0.35\textwidth]{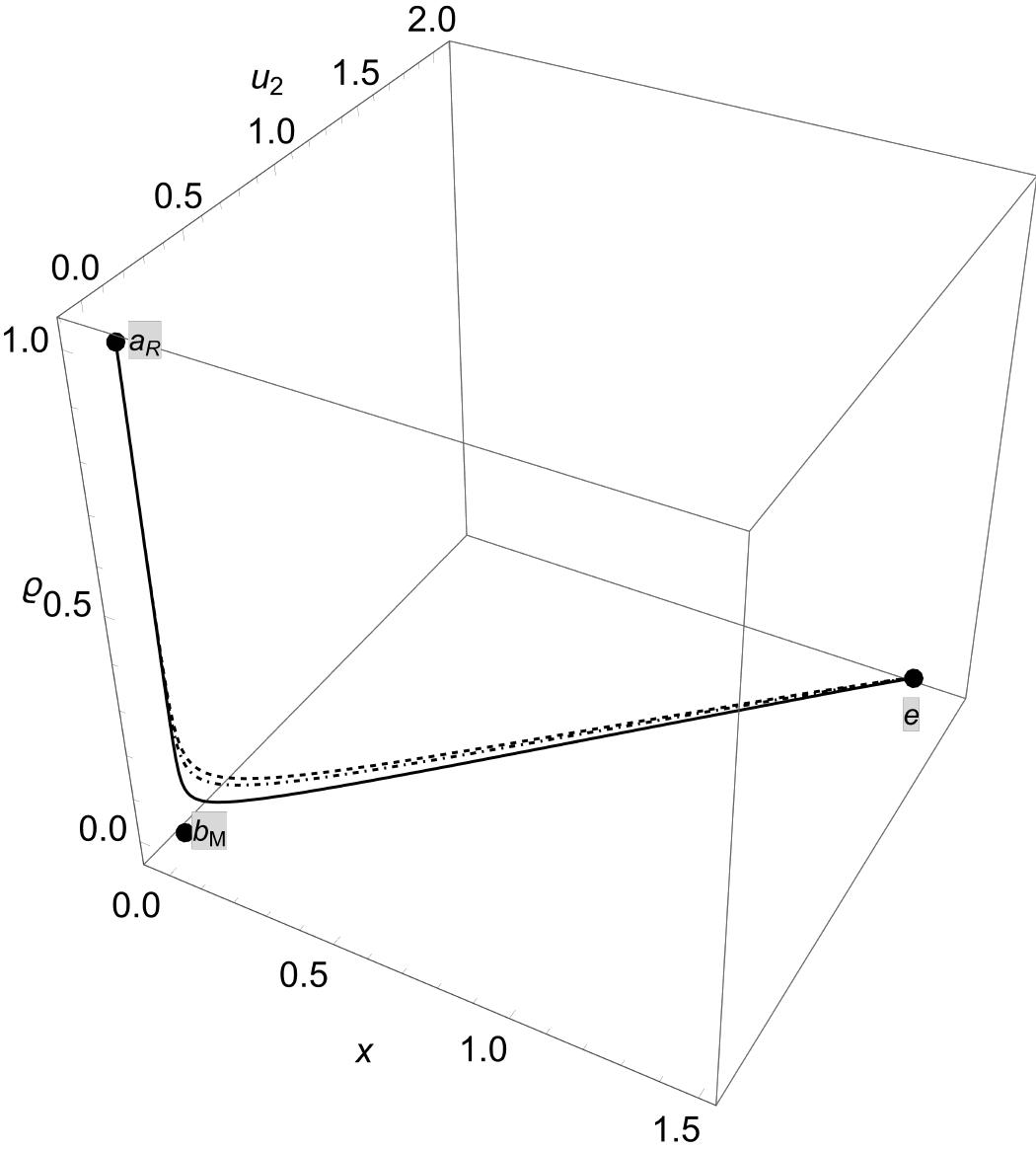}
	\caption{\scriptsize{ Evolution curves in the phase space for $\lambda = 1 $, $\sigma_1 = 1$ and $\sigma_2 = 10^{-3}$. In particular, black solid lines correspond to initial conditions $x_{i} = 1.7 \times 10^{-12}$, $y_{i} = 7.3 \times 10^{-14}$, $u_{1 i} = 5.75 \times 10^{-12}$, $u_{2 i} = 1.7 \times 10^{-12}$ and $\varrho_{i} =  0.99983$. Black dashed lines correspond to initial conditions $x_{i} = 1.7 \times 10^{-8}$, $y_{i} = 7.3 \times 10^{-14}$, $u_{1 i} = 5.75 \times 10^{-11}$, $u_{2 i} = 1.71 \times 10^{-12}$ and $\varrho_{i} =  0.99983$. Black dot-dashed lines correspond to initial conditions $x_{i} = 1.7 \times 10^{-12}$, $y_{i} = 7.3 \times 10^{-14}$, $u_{1 i} = 5.75 \times 10^{-10}$, $u_{2 i} = 1.69 \times 10^{-12}$ and $\varrho_{i} =  0.99983$.}} 
	\label{Figura1cp}
\end{figure}


\begin{figure}[!tbp]
  \centering
  \begin{minipage}[b]{0.45\textwidth}
    \includegraphics[width=\textwidth]{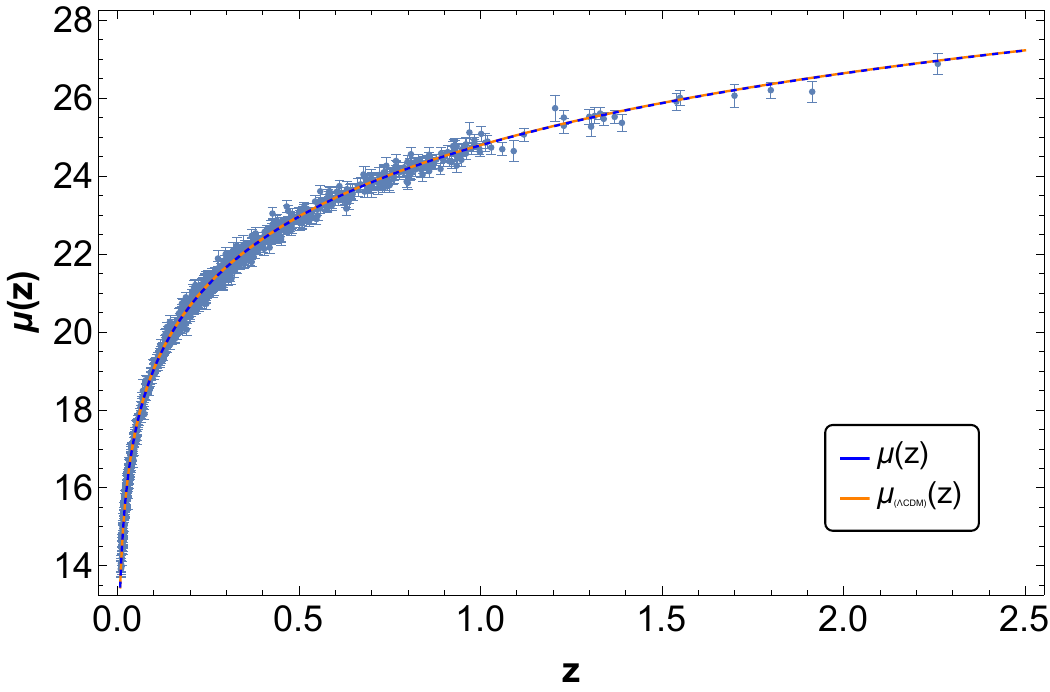}
    \caption{\scriptsize{Distance modulus $\mu(z)$ as a function of the redshift z, contrasted with the Pantheon data set. We used the same initial conditions of FIG.  \ref{Figura1rho} to obtain the solid, dashed, and dot-dashed blue lines.\\}} 
    \label{Figura1mu}
  \end{minipage}
  \hfill
  \begin{minipage}[b]{0.45\textwidth}
    \includegraphics[width=\textwidth, height=4cm]{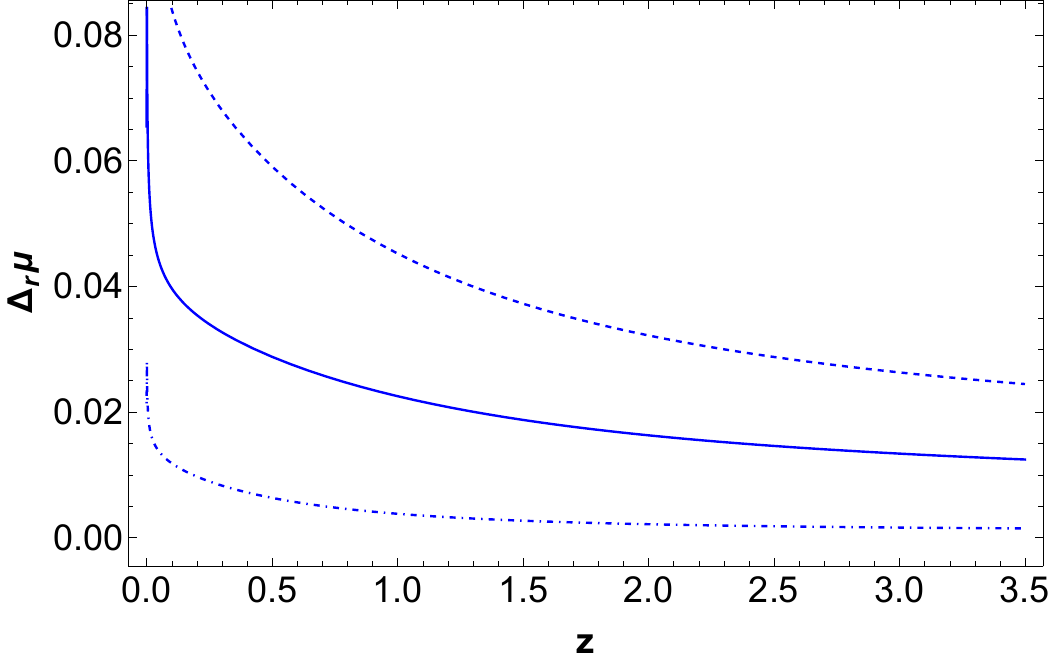}
    \caption{\scriptsize{We show the behavior of relative difference (percentage), $\Delta_r \mu(z) = 100 \times \left| \mu - \mu_{\Lambda CDM} \right|/\mu_{\Lambda CDM}$, with respect to $\Lambda CDM$ model. We used the same initial conditions of FIG.  \ref{Figura1rho} to obtain the solid, dashed, and dot-dashed blue lines.}} 
    \label{Figura1dmu}
  \end{minipage}
\end{figure}
\begin{figure}[!tbp]
  \centering
  \begin{minipage}[b]{0.45\textwidth}
    \includegraphics[width=\textwidth, height=4cm]{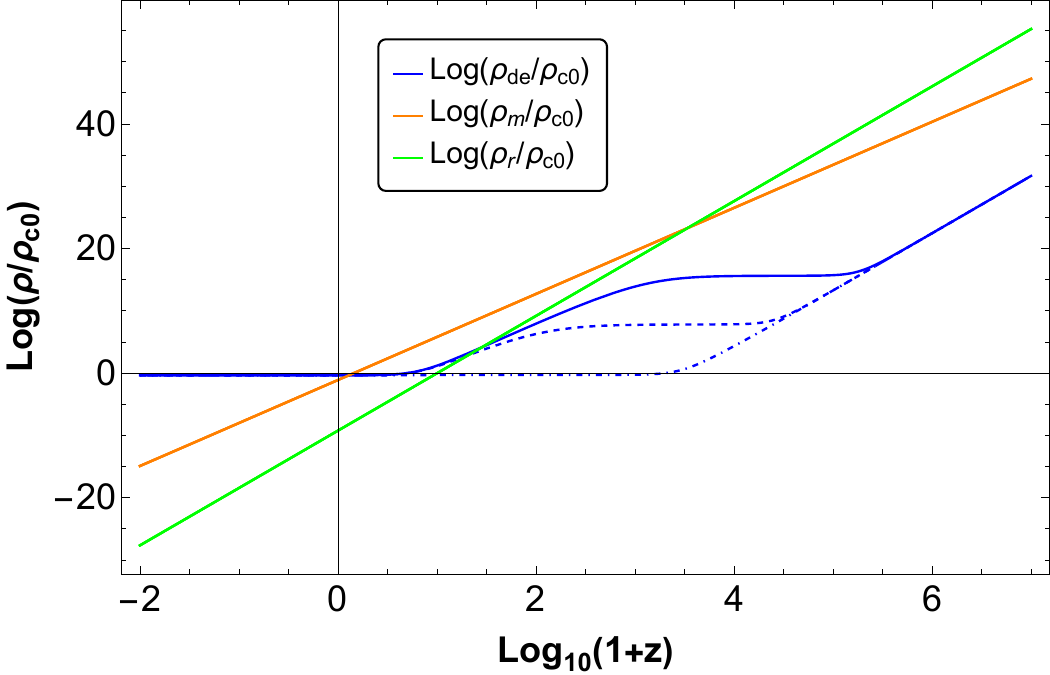}
    \caption{\scriptsize{We present the evolution of the energy density of dark energy, denoted as $\rho_{de}$ (depicted in blue), the energy density of dark matter (including baryons), denoted as $\rho_m$ (shown in orange), and the energy density of radiation, denoted as $\rho_r$ (illustrated in green), as functions of the redshift $z$. The values of the parameters used are $\lambda = 10^{-3}$, $\sigma_1 = 10^{-1}$, and $\sigma_2 = 2.0 \times 10^{1}$. The solid lines represent the cases with initial conditions $x_i = 1.7 \times 10^{-8}$, $y_i = 8.37 \times 10^{-13}$, $u_{1i} = 5.75 \times 10^{-11}$, $u_{2i} = 0.5 \times 10^{-8}$, and $\varrho_i = 0.99983$. On the other hand, the dashed lines correspond to initial conditions $x_i = 1.0 \times 10^{-10}$, $y_i = 8.37 \times 10^{-13}$, $u_{1i} = 5.75 \times 10^{-11}$, $u_{2i} = 1.0 \times 10^{-10}$, and $\varrho_i = 0.99983 $. Lastly, the dot-dashed lines represent the cases with initial conditions $x_i = 5.0 \times 10^{-13}$, $y_i = 8.37 \times 10^{-13}$, $u_{1i} = 5.75 \times 10^{-11}$, $u_{2i} = 5.0 \times 10^{-13}$, and $\varrho_i = 0.99983$.\\
    }} 
    \label{Figura2rho}
  \end{minipage}
  \hfill
  \begin{minipage}[b]{0.45\textwidth}
    \includegraphics[width=\textwidth, height=4cm]{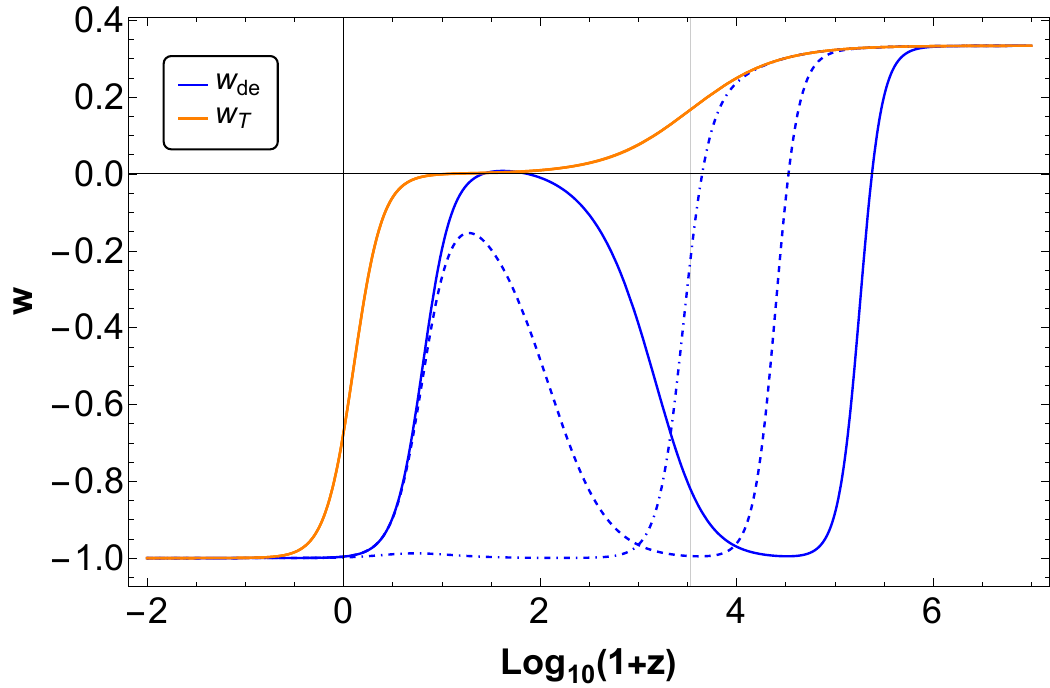}
    \caption{\scriptsize{The equation of state $w_{tot}$ (depicted by the orange line) and the dark energy equation of state $w_{de}$ (represented by the blue line) are plotted as functions of redshift $z$. The solid, dashed, and dot-dashed blue lines correspond to the same initial conditions as shown in FIG.  \ref{Figura2rho}. At $z=0$, we obtain a value of $w_{de}=-0.998427$ for the dark energy equation of state. This value is in agreement with the Planck data value of $w_{de}^{(0)}=-1.028\pm0.032$ as reported in the Ref. \cite{Akrami:2018odb}.}} 
    \label{Figura2omega}
  \end{minipage}
\end{figure}

\begin{figure}[!tbp]
  \centering
  \begin{minipage}[b]{0.45\textwidth}
    \includegraphics[width=\textwidth, height=4cm]{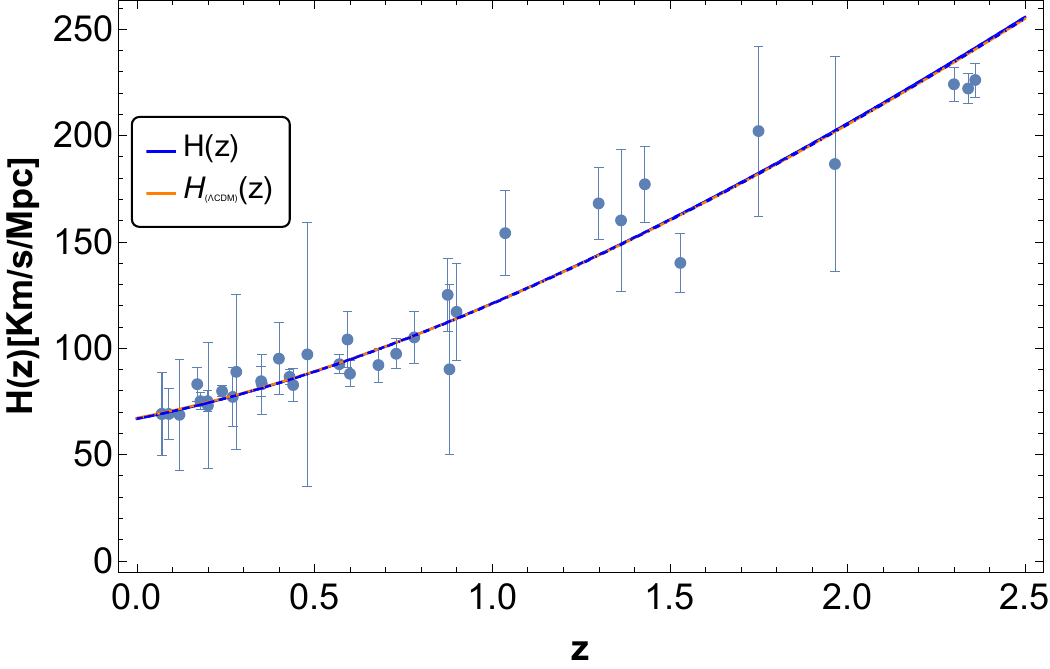}
    \caption{\scriptsize{We depict the evolution of Hubble rate $H(z)$ as a function of red-shift z and H(z) data (blue points) and their $1 \sigma$ confidence interval listed in Table \ref{table:H(z)data}. Where we used the same initial conditions of FIG. \ref{Figura2rho}.\\ \\}} 
    \label{Figura2H}
  \end{minipage}
  \hfill
  \begin{minipage}[b]{0.45\textwidth}
    \includegraphics[width=\textwidth, height=4cm]{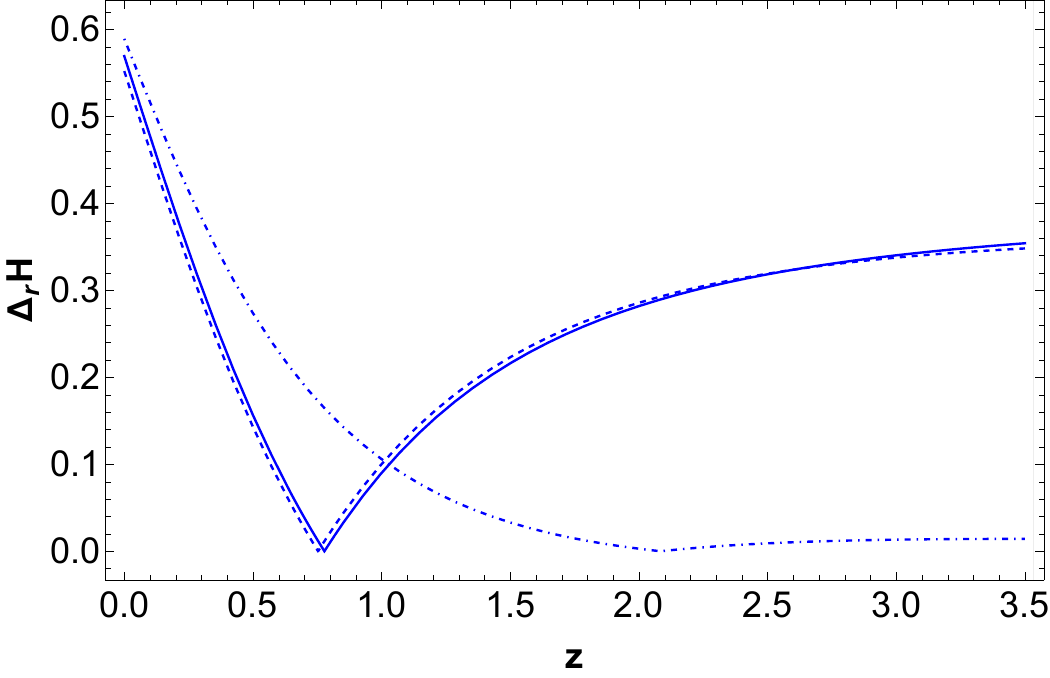}
    \caption{\scriptsize{We show the behavior of relative difference (percentage), $\Delta_r H(z) = 100 \times \left| H - H_{\Lambda CDM} \right|/H_{\Lambda CDM}$, with respect to $\Lambda CDM$ model. We used the same initial conditions of FIG. \ref{Figura2rho} to obtain the solid, dashed, and dot-dashed blue lines.}} 
    \label{Figura2dH}
  \end{minipage}
\end{figure}


\begin{figure}[!tbp]
	\centering
		\includegraphics[width=0.35\textwidth]{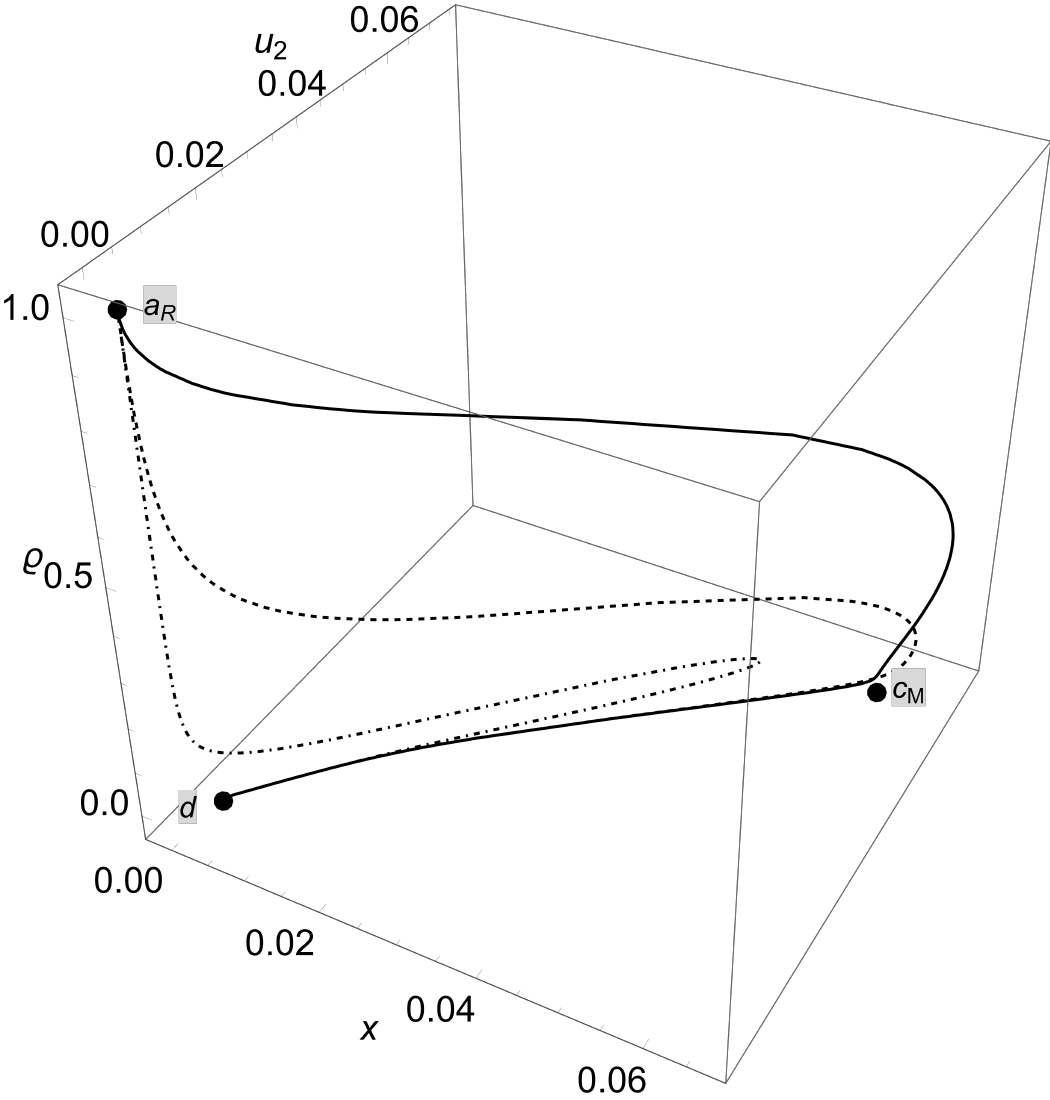}
	\caption{\scriptsize{Evolution curves in the phase space for $\lambda = 10^{-3}$, $\sigma_1 = 10^{-1}$, and $\sigma_2 = 2.0 \times 10^{1}$. The solid lines represent the cases with initial conditions $x_i = 1.7 \times 10^{-8}$, $y_i = 8.37 \times 10^{-13}$, $u_{1i} = 5.75 \times 10^{-11}$, $u_{2i} = 0.5 \times 10^{-8}$, and $\varrho_i = 0.99983$. On the other hand, the dashed lines correspond to initial conditions $x_i = 1.0 \times 10^{-10}$, $y_i = 8.37 \times 10^{-13}$, $u_{1i} = 5.75 \times 10^{-11}$, $u_{2i} = 1.0 \times 10^{-10}$, and $\varrho_i = 0.99983 $. Lastly, the dot-dashed lines represent the cases with initial conditions $x_i = 5.0 \times 10^{-13}$, $y_i = 8.37 \times 10^{-13}$, $u_{1i} = 5.75 \times 10^{-11}$, $u_{2i} = 5.0 \times 10^{-13}$, and $\varrho_i = 0.99983$.}} 
	\label{Figura2cp}
\end{figure}


\begin{figure}[!tbp]
  \centering
  \begin{minipage}[b]{0.45\textwidth}
    \includegraphics[width=\textwidth]{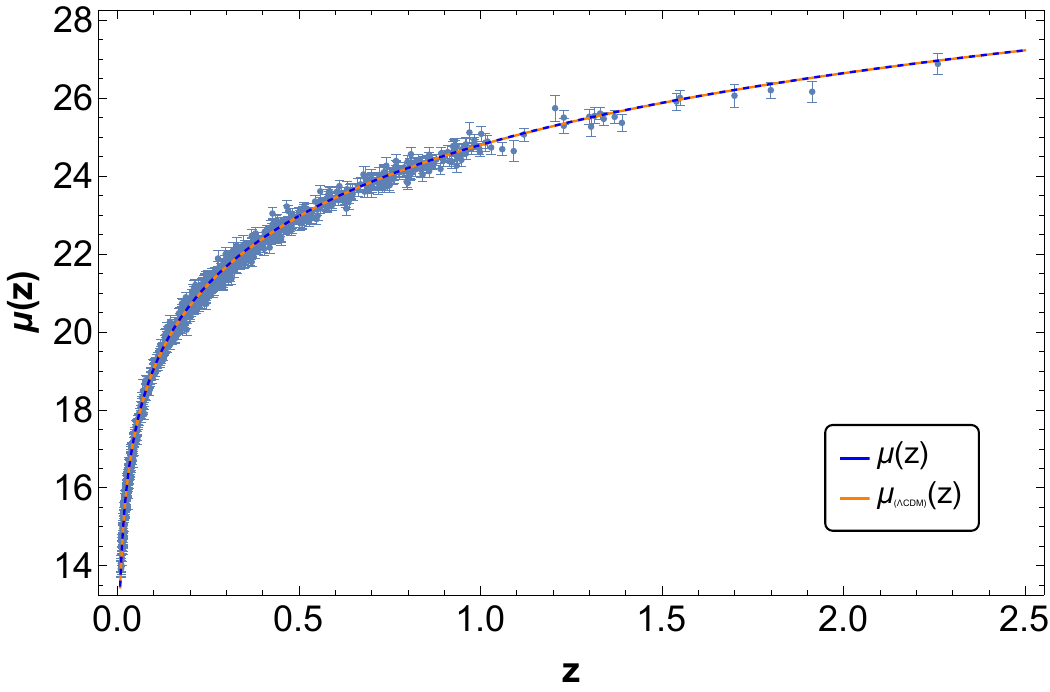}
    \caption{\scriptsize{Distance modulus $\mu(z)$ as a function of the redshift z, contrasted with the Pantheon data set. We used the same initial conditions of FIG. \ref{Figura2rho} to obtain the solid, dashed, and dot-dashed blue lines.\\}} 
    \label{Figura2mu}
  \end{minipage}
  \hfill
  \begin{minipage}[b]{0.45\textwidth}
    \includegraphics[width=\textwidth, height=4cm]{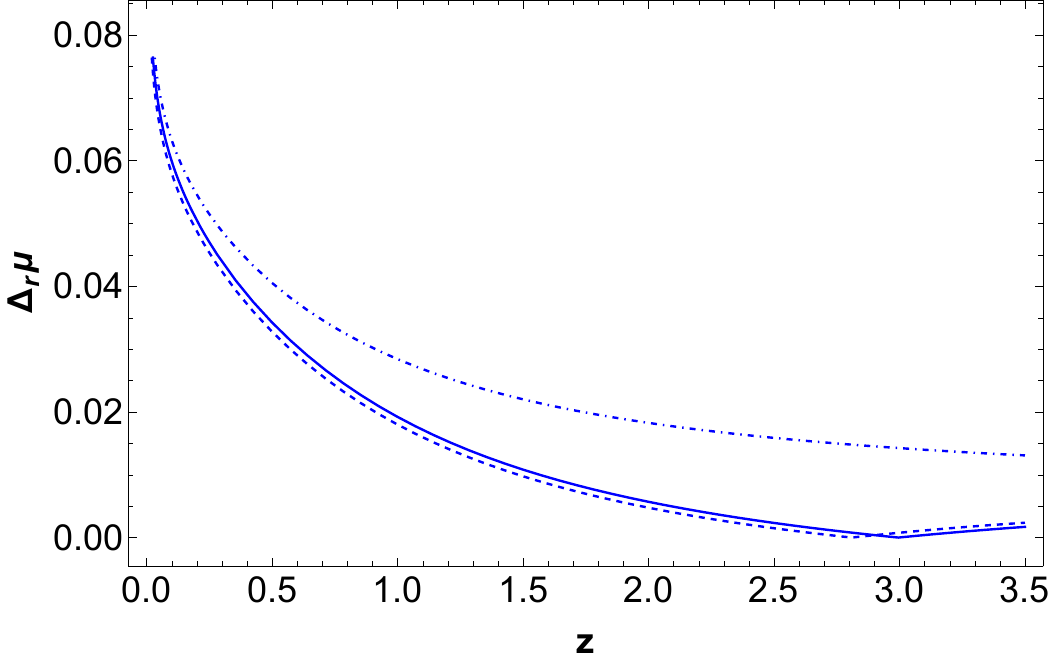}
    \caption{\scriptsize{We show the behavior of relative difference (percentage), $\Delta_r \mu(z) = 100 \times \left| \mu - \mu_{\Lambda CDM} \right|/\mu_{\Lambda CDM}$, with respect to $\Lambda CDM$ model. We used the same initial conditions of FIG. \ref{Figura1rho} to obtain the solid, dashed, and dot-dashed blue lines.}} 
    \label{Figura2dmu}
  \end{minipage}
\end{figure}
\section{Numerical Results}\label{Num_Res}

In this section, we numerically solve the autonomous system, \eqref{dinsyseq1} - \eqref{dinsyseq5} associated with the set of cosmological equations. We investigate the features of our model to explain the current accelerated expansion of the Universe and then the predicted results are contrasted with the latest 
$H(z)$ and Supernovae Ia (SNe Ia) observational data.

\subsection{H(z)}

To analyze the behavior of the Hubble rate and its confidence interval, we will use a collection of $39$ data points obtained by \cite{Farooq:2016zwm,Ryan:2018aif}, which are presented in Table \ref{table:H(z)data} (Appendix \ref{appen_B}) and are expressed in terms of redshift $z$.

Additionally, to perform further comparisons, we make use of the $\Lambda CDM$ model, which provides us a Hubble rate that varies as a function of the redshift as follows
\be
H_{\Lambda CDM}(z)=H_{0}\sqrt{\Omega_{de}^{(0)}+\Omega_{m}^{(0)}(1+z)^3+\Omega_{r}^{(0)}(1+z)^4} .
\ee

In FIGS. \ref{Figura1H} and \ref{Figura2H}, we show the evolution of the Hubble rate as a function of the redshift $z$ for our model. We have used the same values of the parameters and the initial conditions used for plots \ref{Figura1rho} and \ref{Figura2rho}, respectively.  We compared with the corresponding standard results from $\Lambda$CDM and the current observational $H(z)$ data. Thus, we conclude that our results are compatible with the latest observational $H(z)$ data. In FIGS. \ref{FiguradH1} and  \ref{Figura2dH} we depict the behavior of the relative difference, $\Delta_r H(z)$, between the results from our model and $\Lambda$CDM. Thus, we can check that our results are very close to $\Lambda$CDM.

\subsection{Supernovae Ia}

An alternative way of defining distances in an expanding universe is through the properties of the luminosity of a stellar object. The luminosity distance in a flat FLRW universe is given by \cite{Copeland:2006wr}
\be
D_{L}(z)=\frac{1+z}{H_{0}}\int_{0}^{z} \frac{1}{h(z')} dz',
\ee where $h(z)\equiv H(z)/H_{0}$. For analytical solutions of the Hubble rate $H(z)$, we can use this latter integral equation to find $D_{L}(z)$. On the other hand, this equation can also be written in a differential form as follows
\be
\frac{dD_{L}(z)}{dz}-\frac{D_{L}(z)}{1+z}-\frac{1+z}{H(z)}=0. 
\ee This latter equation can be useful to integrate $D_{L}(z)$ when we do not have an analytical solution for $H(z)$.

The direct evidence for dark energy is supported by the observation of the luminosity distance of high redshift supernovae. Related to the same definition of the luminosity distance, the difference between the apparent magnitude $m$ of the source and its absolute magnitude $M$ can be computed as 
\be
\mu(z)\equiv m-M =5 Log_{10}\left(\frac{D_{L}(z)}{M_{pc}}\right)+25, 
\ee which is called distance modulus. The numerical factor comes from the conventional definitions of $m$ and $M$ in astronomy \cite{Copeland:2006wr,amendola2010dark}. 

In FIGS. \ref{Figura1mu} and \ref{Figura2mu}, we show the behavior of $\mu(z)$ for our model using the same parameter values and initial conditions used in plots \ref{Figura1rho} and \ref{Figura2rho}, respectively. We have contrasted our theoretical results with the latest SNe Ia observational data.  
Specifically, we focus on the SNe Ia data in the Pantheon sample \cite{Pan-STARRS1:2017jku}, which comprises 1048 supernova data points covering the redshift range $0.01 < z < 2.3$.\footnote{Data available online in the GitHub repository \\ \url{https://github.com/dscolnic/Pantheon}.} Furthermore, in FIGS. \ref{Figura1dmu} and \ref{Figura2dmu}, we have depicted the behavior of the relative difference between our model and $\Lambda$CDM. Thus, we can check that our results are compatible with the current SNe Ia observational data, and they are very close to $\Lambda$CDM results. 

It is important to note that our main goal was to corroborate that our theoretical results are compatible with the current SNe Ia observational data. On the other hand, a different point of view can be followed when using observational data to constrain the parameter ranges for the model \cite{Copeland:2006wr, Aghanim:2018eyx}. This latter approach lies beyond the scope of the present work, and so it is left for a separate project.

\section{Conclusions}\label{conclusion_f}
We studied the cosmological dynamics of dark energy in a torsion-coupled dilatonic ghost condensate model. In this context, torsion is defined from the Weitzenb\"{o}ck connection of teleparallel gravity \cite{Aldrovandi-Pereira-book,JGPereira2,Arcos:2005ec}. This is a flat connection with non-vanishing torsion in the presence of gravitation. We calculated the set of cosmological equations and then the corresponding autonomous system. Thus, we performed a detailed phase space analysis by obtaining all the critical points and their stability conditions.  Furthermore, we have compared our results with the latest $H(z)$ and Supernovae Ia observational data. 

We found critical points describing the current accelerated epoch, as is the case of de Sitter attractors \cite{amendola2010dark}. Thus, independently of the initial conditions, and provided that they are close to the critical point, the system will fall into a dark energy dominated epoch with accelerated expansion. Also, we have shown that this accelerated epoch at late times can be connected with the standard radiation and matter eras. Furthermore, we have proved the existence of a matter scaling era. So, we have found the corresponding conditions for the model parameters such that the system can successfully exit from this scaling regime towards a dark energy dominated attractor point with acceleration. It is worth noting that the presence of scaling solutions represents a highly intriguing feature in any cosmological model \cite{amendola2010dark}. These kinds of solutions allow us to naturally incorporate early dark energy, providing an additional phenomenological behavior that is absent in the $\Lambda$CDM model. Thus the scaling solutions can be used as a mechanism to alleviate the energy scale problem of the $\Lambda$CDM model due to the large energy gap between the critical energy density of the Universe today and the typical energy scales of particle physics \cite{Albuquerque:2018ymr,Ohashi:2009xw}. Therefore, a model that admits the existence of scaling solutions with a dark energy component during the early universe can lead to new imprints in early-time physics, allowing to contrast its predictions with the current observational data \cite{Koivisto:2008xf,Doran:2006kp}.\\

\begin{acknowledgments}
M. Gonzalez-Espinoza acknowledges the financial support of FONDECYT de Postdoctorado, N° 3230801. G. Otalora acknowledges Dirección de Investigación, Postgrado y Transferencia Tecnológica de la
Universidad de Tarapacá for financial support through
Proyecto UTA Mayor 4731-23. J. Saavedra acknowledges the financial support of Fondecyt Grant 1220065.
\end{acknowledgments}


\bibliography{bio} 



\begin{appendix}

\section{Stability of quantum fluctuations in the presence of dilatonic ghost condensate field}\label{appen_A}

Starting with a general Lagrangian density $p(\phi, X)$, upon expanding it to the second order in $\delta\phi$, we can easily derive the corresponding Lagrangian and Hamiltonian for the fluctuations \cite{Piazza:2004df,Copeland:2006wr}. The Hamiltonian can be expressed as follows 
\begin{equation}
\mathcal{H}=\left(p_{, X}+2 X p_{, X X}\right) \frac{(\delta \dot{\phi})^2}{2}+p_{, X} \frac{(\nabla \delta \phi)^2}{2}-p_{, \phi \phi} \frac{(\delta \phi)^2}{2},
\end{equation}
where $ p_{,X}  \equiv \partial p / \partial X$. And the Hamiltonian remains positive as long as the following conditions are satisfied
\begin{equation}
\begin{aligned}
& \xi_1 \equiv p_{, X}+2 X p_{, X X} \geq 0, \quad \xi_2 \equiv p_{, X} \geq 0 ,\\
& \xi_3 \equiv-p_{, \phi \phi} \geq 0 .
\end{aligned}
\end{equation}

To ensure the stability of quantum fluctuations, it is necessary that both $\xi_1 \geq 0$ and $\xi_2 \geq 0$ \cite{Copeland:2006wr}.
It can be expected that these conditions will not be modified by a nonminimal coupling if there is no direct coupling between torsion and the kinetic term of the scalar field \cite{Lopez:2021agu,Gonzalez-Espinoza:2020azh}. 
\begin{figure}[!b]
  \centering
  \begin{minipage}[b]{0.45\textwidth}
    \includegraphics[width=\textwidth, height=3.5cm]{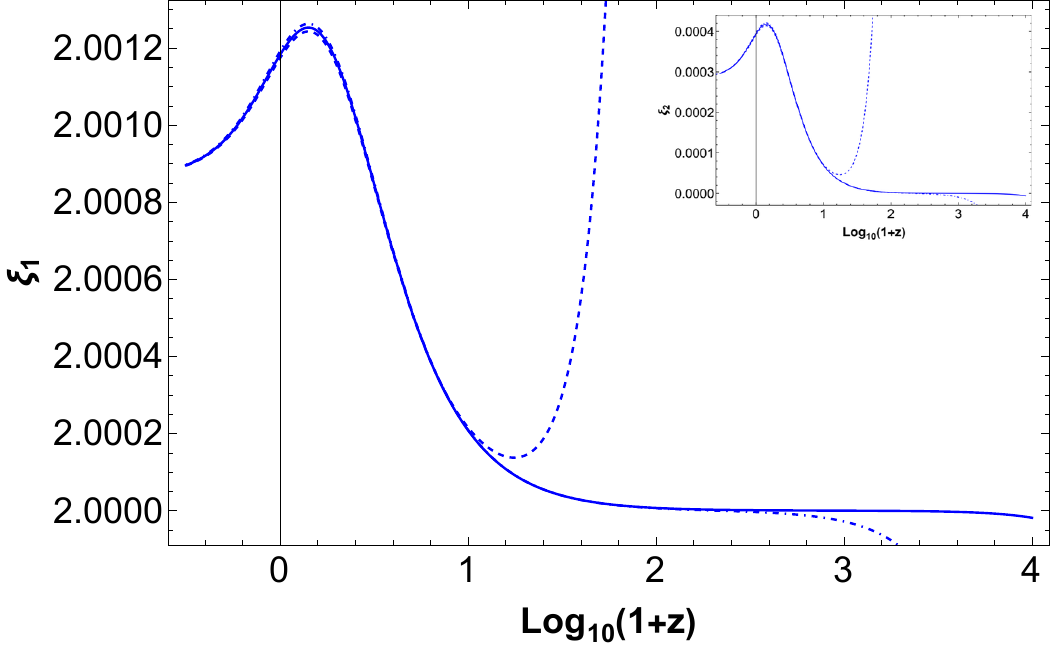}
    \caption{\scriptsize{We show the behavior of $\xi_1$ and $\xi_2$ (small plot). Where we used the same initial conditions of FIG. \ref{Figura1rho} to obtain the solid, dashed, and dot-dashed blue lines.\\}} 
    \label{Figura1xi}
  \end{minipage}
  \hfill
  \begin{minipage}[b]{0.45\textwidth}
    \includegraphics[width=\textwidth, height=3.5cm]{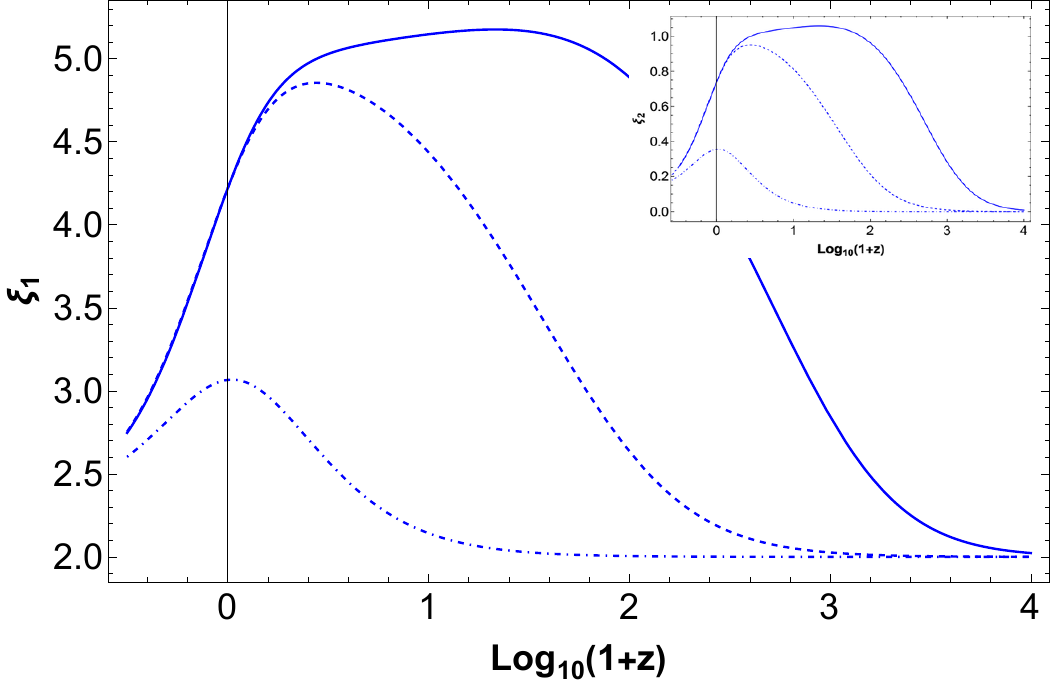}
    \caption{\scriptsize{We show the behavior of $\xi_1$ and $\xi_2$ (small plot). Where we used the same initial conditions of FIG. \ref{Figura2rho} to obtain the solid, dashed, and dot-dashed blue lines.}} 
    \label{Figura2xi}
  \end{minipage}
\end{figure}
Therefore, we analyze our case, $p = - X - V(\phi) + F_2 X^2$,
\begin{equation}
\begin{aligned}
& \xi_1 = - 1 + 6 X F_2 = -1 + 6 \dfrac{x^2}{u^2_2}, \\
& \xi_2 = - 1 + 2 X F_2 = -1 + 2 \dfrac{x^2}{u^2_2},
\end{aligned}
\end{equation}
obtaining stability for quantum fluctuations of studied cases, FIG. \ref{Figura1xi} and FIG. \ref{Figura2xi}.

\begin{table}[!b]
\caption{Hubble's parameter vs. redshift \& scale factor.}
\label{table:H(z)data}
\renewcommand{\tabcolsep}{0.7pc} 
\renewcommand{\arraystretch}{0.7} 
\begin{tabular}{@{}lllll}
\hline \hline
  $\;\; z$    &  $ H(z) \;$ ($\frac{km/s}{\text{Mpc}}$ ) &  Ref. \\
\hline
$0.07$      & $ \; \qquad 69     \pm 19.6 $      &   \cite{zhang2014} \\
$0.09$      & $ \; \qquad 69     \pm 12 $      & \cite{simon2005} \\
$0.100$     & $ \; \qquad 69     \pm 12 $      & \cite{simon2005} \\
$0.120$     & $ \; \qquad 68.6     \pm 26.2$       & \cite{zhang2014} \\
$0.170$     & $ \; \qquad 83     \pm 8$       & \cite{simon2005} \\
$0.179$     & $ \; \qquad 75     \pm 4$       & \cite{moresco2012} \\
$0.199$     & $ \; \qquad 75     \pm 5$        & \cite{moresco2012} \\
$0.200$     & $ \; \qquad 72.9     \pm 29.6$        &  \cite{zhang2014} \\
$0.270$     & $ \; \qquad 77     \pm 14$      & \cite{simon2005} \\
$0.280$     & $ \; \qquad 88.8     \pm 36.6$      & \cite{zhang2014} \\
$0.320$     & $ \; \qquad 79.2   \pm 5.6$     & \cite{cuesta2016}\\
$0.352$     & $ \; \qquad 83     \pm 14$      & \cite{moresco2012} \\
$0.3802$    & $ \; \qquad 83     \pm 13.5$      & \cite{moresco2012} \\
$0.400$     & $ \; \qquad 95     \pm 17$      & \cite{simon2005} \\
$0.4004$    & $ \; \qquad 77     \pm 10.2$      & \cite{moresco2012} \\
$0.4247$    & $ \; \qquad 87.1     \pm 11.2$      & \cite{moresco2012} \\
$0.440$     & $ \; \qquad 82.6   \pm 7.8$     & \cite{blake2012} \\
$0.4497$    & $ \; \qquad 92.8   \pm 12.9$     & \cite{moresco2012} \\
$0.470$     & $ \; \qquad 89   \pm 50$     & \cite{ratsim} \\
$0.4783$    & $ \; \qquad 80.9   \pm 9$     & \cite{moresco2012} \\
$0.480$     & $ \; \qquad 97     \pm 62$      & \cite{stern2010} \\
$0.570$     & $ \; \qquad 100.3  \pm 3.7$     & \cite{cuesta2016} \\
$0.593$     & $ \; \qquad  104   \pm 13$      & \cite{moresco2012} \\
$0.600$     & $ \; \qquad 87.9   \pm 6.1$     & \cite{blake2012} \\
$0.680$     & $ \; \qquad 92     \pm 8$       & \cite{moresco2012} \\
$0.730$     & $ \; \qquad 97.3   \pm 7 $      & \cite{blake2012} \\
$0.781$     & $ \; \qquad 105    \pm 12$      & \cite{moresco2012} \\
$0.875$     & $ \; \qquad 125    \pm 17$      & \cite{moresco2012} \\
$0.880$     & $ \; \qquad 90     \pm 40$      & \cite{stern2010} \\
$0.900$     & $ \; \qquad 117    \pm 23$      & \cite{simon2005} \\
$1.037$     & $ \; \qquad 154    \pm 20 $     & \cite{moresco2012} \\
$1.300$     & $ \; \qquad 168    \pm 17 $     & \cite{simon2005} \\
$1.363$     & $ \; \qquad 160    \pm 33.6$    & \cite{moresco2015}\\
$1.430$     & $ \; \qquad 177    \pm 18$      & \cite{simon2005}\\
$1.530$     & $ \; \qquad 140    \pm 14$      & \cite{simon2005}\\
$1.750$     & $ \; \qquad 202    \pm 40$      & \cite{simon2005}\\
$1.965$     & $ \; \qquad 186.5  \pm 50.4$    & \cite{moresco2015}\\
$2.340$     & $ \; \qquad 222    \pm 7 $      & \cite{delubac2014}\\
$2.360$     & $ \; \qquad 226    \pm  8$      & \cite{font-ribera2014}\\
\hline
\end{tabular}\\
 \end{table}
\section{Hubble's parameter data}\label{appen_B}

In this appendix, we present Hubble's parameter data for $0.01 < z < 2.360$:

\end{appendix}

\end{document}